\documentclass[superscriptaddress,twocolumn,showpacs,prl,floatfix]{revtex4}

\usepackage{color}
\usepackage{tabularx}
\usepackage{epsfig}
\usepackage{amsmath}

\usepackage{graphicx}

\renewcommand{\log}{\ln}
\newcommand{\dd}{\mathrm{d}}

\begin{document}

\title{Link overlaps at Criticality and Universality in Ising Spin Glasses}

\author{P. H.~Lundow} 
\affiliation {Department of Theoretical Physics,
  Kungliga Tekniska h\"ogskolan, SE-106 91 Stockholm, Sweden}

\author{I. A.~Campbell}
\affiliation{Laboratoire Charles Coulomb,
  Universit\'e Montpellier II, 34095 Montpellier, France}

\begin{abstract}
Extensive simulations are made of link and spin overlaps in four and
five dimensional Ising Spin Glasses (ISGs). Moments and moment ratios of
the mean link overlap distributions (the variance, the kurtosis and
the skewness) show clear critical behavior around the ISG ordering
temperature. The link overlap measurements can be used to identify the
ISG transition accurately; the link overlap is often a more efficient
tool in this context than the spin overlap because the link overlap
inter-sample variability is much weaker. Once the transition
temperature is accurately established, critical exponents can be
readily estimated by extrapolating measurements made in the
thermodynamic limit regime. The data show that the bimodal and
Gaussian spin glass susceptibility exponents $\gamma$ are different
from each other, both in dimension $5$ and in dimension $4$. Hence ISG
critical exponents are not universal in a given dimension, but depend
on the form of the interaction distribution.
\end{abstract}


\maketitle

\section{Introduction}

We have studied the equilibrium link and spin overlap distributions
(defined below, Eqs.~\eqref{qtdef} and \eqref{qltdef}) in some detail for
Ising Spin Glasses (ISGs) on [hyper]cubic lattices with bimodal and
Gaussian near neighbor interaction distributions in dimension five and
with bimodal near neighbor interactions in dimension four.

The Hamiltonian is as usual
\begin{equation}
  \mathcal{H}= - \sum_{ij}J_{ij}S_{i}S_{j}
  \label{ham}
\end{equation}
with the near neighbor symmetric bimodal ($\pm J$) or Gaussian
interaction distributions normalized to $\langle
J_{ij}^2\rangle=1$. Throughout we will quote inverse temperatures
$\beta = 1/T$. 

The link overlap parameter \cite{caracciolo:90} in ISG numerical
simulations is the bond analogue of the intensively studied spin
overlap. In both cases two replicas (copies) $A$ and $B$ of the same
physical system are first generated and equilibrated; updating is then
continued and the "overlaps" between the two replicas are recorded
over long time intervals. The spin overlap at any instant $t$
corresponds to the fraction $q(t)$ of spins in $A$ and $B$ having the
same orientation (both up or both down), and the normalized overall
distribution over time is written $P(q)$. The link overlap corresponds
to the fraction $q_{\ell}(t)$ of links (or bonds or edges) between
spins which are either both satisfied or both dissatisfied in the two
replicas; the normalized overall distribution over time is written
$Q(q_{\ell})$.  The explicit definitions are for the spin overlap
\begin{equation}
  q(t)=\frac{1}{N}\,\sum_{i=1}^{N} S_{i}^{A}(t)S_{i}^{B}(t)
  \label{qtdef}
\end{equation}
and for the link overlap
\begin{equation}
  q_{\ell}(t)=\frac{1}{N_{\ell}}\sum_{ij}S_{i}^{A}(t)S_{j}^{A}(t)S_{i}^{B}(t)S_{j}^{B}(t)
  \label{qltdef}
\end{equation}
where $N$ is the number of spins per sample and $N_{\ell}$ the number of
links; near neighbor spins $i$ and $j$ are linked, as denoted by $ij$.
We will indicate means taken over time for a given sample by
$\langle\cdots\rangle$ and means over sets of samples by
$\lbrack\cdots\rbrack$.  The physical distinction between the
information obtained from $P(q)$ and $Q(q_{\ell})$ is frequently
illustrated in terms of a low temperature domain picture
\cite{bokil:00}.  "Overlap equivalence" has been proved in
\cite{contucci:06}.

The critical behavior of the link overlaps in ISGs has not been
studied before, as far as we are aware.  It turns out that the moments
and moment ratios of the link overlap distributions have
characteristic forms as functions of temperature around $\beta_c$ and
that these data can be used to supplement and improve on information
from spin overlap measurements.

It is widely considered to be self-evident that just as the standard
universality rules hold exactly at ferromagnetic ordering transitions,
they should hold also for ISG transitions. Early exponent estimates
were erratic (see a summary in Ref.~\cite{katzgraber:06}).
Recent careful comparisons between the critical exponent estimates for
bimodal and Gaussian interaction distribution ISGs have indeed
concluded that both in $3$d \cite{katzgraber:06, hasenbusch:08} and in
$4$d \cite{jorg:08} the exponents for the two systems are the same to
within numerical precision. High Temperature Series Expansion (HTSE)
analyses also concluded that the estimates for the exponent $\gamma$
for different ISGs were compatible with universality to within the
precision of the method \cite{daboul:04}. However, the application of
the standard universality rules to ISGs has been questioned on the
basis of dynamic simulations \cite{bernardi:96,henkel:05} and it is
relevant that the critical exponents in $3$d Heisenberg spin glasses
have been shown experimentally to depend on the strength of the
Dzyaloshinsky-Moriya anisotropy \cite{campbell:10}, and so are not
universal.

The present simulation results concern ISGs in the high dimensions
$d=5$ and $d=4$, where independent information from HTSE can be used
in conjunction with the numerical data. Once accurate values of
ordering temperatures have been obtained using information from a
combination of spin overlap, link overlap and HTSE data, the critical
exponent $\gamma$ can readily be estimated from the temperature
variation of the spin glass susceptibility in the paramagnetic
state. The data presented below show that in the same dimension,
bimodal and Gaussian ISGs have different values for $\gamma$, so the
standard simple universality rules are not obeyed.


\section{Spin overlap parameters}

The spin glass susceptibility is defined by
\begin{equation}
  \chi(\beta,L) = N\,\lbrack\langle q(t)^2\rangle\rbrack
\end{equation}
where $q(t)$ is the spin overlap Eqn.~\eqref{qtdef}.  The standard Binder
cumulant criterion which is widely used to estimate the ordering
temperature $\beta_c$ in ISGs consists of the observation of
intersections of mean spin overlap kurtosis curves as functions of
temperature for different sample sizes $L$.  We use "$P$ kurtosis"
Eqn.~\eqref{defPk} to specify the kurtosis of the spin overlap
distribution to distinguish it from the kurtosis of the link overlap
distribution, the "$Q$ kurtosis" Eqn.~\eqref{defQk}. A mean $P$ kurtosis can
be defined either as
\begin{equation}
  P_{k}(\beta,L) = 
  \left\lbrack
  \frac{
    \langle q^4\rangle
  }{
    \langle q^2\rangle^2
  }\right\rbrack
  \label{defPk}
\end{equation}
which we will use here, or alternatively
\begin{equation}
  P_{km}(\beta,L) = 
  \frac{
    \lbrack\langle q^4\rangle\rbrack
  }{
    \lbrack\langle q^2\rangle^2\rbrack
  }
  \label{defPkm}
\end{equation}
which is the definition more frequently used. $P$ kurtosis data are
generally expressed in terms of the Binder cumulant
$g(\beta,L)=\lbrack 3-P_{km}(\beta,L)\rbrack/2$.  One drawback to this
procedure for estimating $\beta_c$ is that the inter-sample
variability of $\langle q^2\rangle$ and {\it a fortiori} of $\langle
q^4 \rangle$ are strong at $\beta_c$ in ISGs. The normalized variation
of the ISG susceptibility (the non-self-averaging parameter)
\begin{equation}
  A = U_{22}= \lbrack\langle q^{2}\rangle^{2}\rbrack/\lbrack\langle q^{2}\rangle\rbrack^{2} - 1
  \label{defA}
\end{equation}
is typically about $0.20$ at $\beta_c$
\cite{hukushima:00,palassini:03,hasenbusch:08}, and results on large
numbers of samples must be recorded at each size to overcome
statistical fluctuations in the mean $P$ kurtosis. There are in general
finite size corrections so
\begin{equation}
  P_k(\beta_c,L)= P_k(\beta_c,\infty) (1+c_{k}L^{-\omega})
  \label{defcorr}
\end{equation}
with prefactor $c_{k}$ and exponent $\omega$ which are {\it a priori}
unknown. Delicate extrapolation to infinite $L$ is required to
estimate the thermodynamic limit critical temperature. To obtain the
intersection point of curves $P_k(\beta,L)$ and $P_k(\beta,L^{'})$ it
is necessary to equilibrate samples up to the larger size $L^{'}$
while the position of the intersection point is affected mainly by the
larger finite size correction of the smaller size $L$.

As well as higher order moment ratios, there are other dimensionless
parameters of the spin overlap distributions which also can be
studied, if the one-sided distributions of the absolute value of the
spin overlap $P_{\mathrm{abs}}(|q|)$ are recorded.  These include the
second moment ratio
\begin{equation}
  P_{W} = 
  \left\lbrack
  \frac{
    \langle q^2\rangle
  }{
    \langle|q|\rangle^2
  }\right\rbrack,
  \label{defPW}
\end{equation}
the mean skewness of the absolute spin overlap distribution
\begin{equation}
  P_{\mathrm{absskew}} = 
  \left\lbrack
  \frac{
    \langle(|q|-\langle|q|\rangle)^3\rangle
  }{
    \langle(|q|-\langle|q|\rangle)^2\rangle^{3/2}
  }
  \right\rbrack,
  \label{defPabskew}
\end{equation}
and the mean kurtosis of the absolute spin overlap distribution
\begin{equation}
  P_{\mathrm{abskurt}}= 
  \left\lbrack
  \frac{
    \langle(|q|-\langle|q|\rangle)^4\rangle
  }{
    \langle(|q|-\langle|q|\rangle)^2\rangle^2
  }
  \right\rbrack
  \label{defPabk}
\end{equation}

It turns out that the first two moment ratio parameters are of the
standard phenomenological coupling form; at $\beta =0$ each takes up a
standard value corresponding to that of a one-sided Gaussian, and then
decreases towards a low value at high $\beta$. As functions of $\beta$
at fixed $L$, at $\beta_c$ to leading order the curves go through a
size-independent critical value and have a maximum slope
$[\dd P(\beta,L)/\dd\beta]_{c}$ whose value increases as $L^{1/\nu}$. The
parameters generally show finite size corrections. On the other hand
data on the novel parameter $P_{\mathrm{abskurt}}(\beta,L)$ defined by
Eqn.~\eqref{defPabk} shows a deep dip as a function of temperature for
each $L$. With increasing $L$ the dip narrows and the position of the
minimum is tending towards $\beta_c$ subject to a weak finite size
correction.  $P_{\mathrm{abskurt}}(\beta,L)$ behaves in quite
different ways in the ISGs and in the ferromagnet
\cite{lundow:12}. This spin overlap based parameter is obviously a
useful supplementary measurement for estimating critical temperatures.

We will not discuss here the correlation length ratio $\xi(\beta,L)/L$
which is a widely used phenomenological coupling parameter independent
of the spin overlap distribution.

\section{Link overlaps}

Turning to the link overlaps, for the Gaussian ISG it has been shown
\cite{katzgraber:01} that in equilibrium
\begin{equation}
  R(\beta) =  
  \frac{
    \lbrack\langle q_{\ell}(L,\beta)\rangle\rbrack
  }{
    (1-\lbrack\langle U(L,\beta) \rangle\rbrack/\beta)
  } = 1
  \label{Rql}
\end{equation}
where $\lbrack\langle U\rangle(\beta,L)\rbrack$ is the mean energy
per bond. The $5$d Gaussian ISG data presented here satisfy this
equilibrium condition over the full temperature range used; the
bimodal samples equilibrated faster than the Gaussian ones and were
equilibrated for as long times so we will consider that for present
purposes effective equilibration has usually been reached. However, the
data show that the condition Eqn.~\eqref{Rql} is necessary but is not
stringent enough to guarantee true equilibration. A stricter and more
general condition, which can be applied whatever the interaction
distribution, is that all spin overlap or link overlap parameters for
each individual sample should vary smoothly with temperature. By
inspection of the individual sample data sets it can be seen if and
when the equilibrium condition begins to break down as the temperature
is lowered. This test shows that for the same $L$ some samples
equilibrate more easily than others, as noted in Ref.~\cite{janus:10}.
What has not been remarked on is that "simple" samples, where the spin
overlap distribution is tending to two pure peaks beyond the ordering
transition and $\langle q_{\ell}\rangle$ is high, equilibrate more
easily than "complex" samples, those for which the spin overlap
distribution remains multi-peaked and $\langle q_{\ell}\rangle$ is low
even at low temperatures.

For the symmetric bimodal ISG there are simple rules on the mean link
overlap $\langle q_{\ell}\rangle$.  If $p_{s}(\beta)$ is the
probability that a bond is satisfied, by definition
\begin{equation}
  U(\beta) \equiv 2p_{s}(\beta) - 1
\end{equation}
and for symmetric interaction distributions where the Nishimori point
is at $\beta=0$ (uncorrelated satisfied bond positions) a strict lower
limit on $\langle q_{\ell}\rangle$ is given by
\begin{equation}
  \lbrack\langle q_{\ell}(L,\beta)\rangle\rbrack \geq p_{s}^2 +
  (1-p_{s})^2 - 2p_{s}(1-p_{s}) \equiv \lbrack
    U(L,\beta)^2\rbrack
\end{equation}
In the high temperature limit $|U|(\beta) \to \tanh(\beta)$ so
$R(\beta) \to 3$.  As $\beta$ increases $R(\beta)$ drops and appears
to tend gradually towards $1$.

For a pure near neighbor ferromagnet (so with translational
invariance) $\lbrack\langle q_{\ell}\rangle(\beta,L)\rbrack / \lbrack
U(\beta,L)^2\rbrack =1$ at all temperatures \cite{lundow:12}. For the
bimodal ISG this ratio is equal to $1$ for small $\beta$ but then
gradually grows as $\beta$ increases and certain bonds become
preferentially satisfied.

Certain moments and moment ratios for the link overlap distributions
$Q(q_{\ell})$ show characteristic critical behavior in the ISGs, as
they do in a ferromagnet \cite{lundow:12}.  Link overlap data for a
given $\beta$ and $L$ can be recorded for virtually no extra
computational cost in a simulation designed for spin overlap
measurements, while the inter-sample variability of the link overlap
distributions at $\beta_c$ is considerably weaker than that of the
spin overlap distributions. This implies that measurements on mean
link overlap values require far fewer samples than those on mean spin
overlap values, or alternatively that with the same number of samples
the mean link overlap measurements are more precise than the mean spin
overlap measurements. Thus link overlap based data are more efficient
for obtaining accurate estimates of critical temperatures, and so of
critical exponents, than are spin overlap data.

As well as the mean link overlap $\lbrack\langle
q_{\ell}\rangle\rbrack$, and the variance of the link overlap
\begin{equation}
  Q_{\mathrm{var}} = N_{\ell}\, \left\lbrack\langle (q_{\ell}- \langle q_{\ell} \rangle)^2 \rangle\right\rbrack
  \label{defQvar}
\end{equation}
we have recorded three dimensionless moment ratios for each sample and
their averages over all samples.  These are the $Q$ kurtosis
\begin{equation}
  Q_{k}(\beta,L) =
  \left\lbrack
  \frac{
    \left\langle\left(q_{\ell}-\langle q_{\ell}\rangle\right)^4\right\rangle
  }{
    \left\langle\left(q_{\ell}-\langle q_{\ell}\rangle\right)^2\right\rangle^2
  }\right\rbrack,
  \label{defQk}
\end{equation}
the $Q$ skewness
\begin{equation}
  Q_{s}(\beta,L) =
  \left\lbrack
  \frac{
    \left\langle\left(q_{\ell}-\langle q_{\ell}\rangle\right)^3\right\rangle
  }{
    \left\langle\left(q_{\ell}-\langle q_{\ell}\rangle\right)^2\right\rangle^{3/2}
  }\right\rbrack,
  \label{defQsk}
\end{equation}
and $Q_{w}$ which is the mean squared signal to noise ratio, or the
mean of the inverse square of the coefficient of variation. This has a
clumsy name but a simple definition:
\begin{equation}
  Q_{w} =  \left\lbrack
  \frac{
    \langle q_{\ell}\rangle^2
  }{
    N_{\ell} \langle(q_{\ell}- \langle q_{\ell}\rangle)^2\rangle
  }\right\rbrack
  \label{defQw}
\end{equation}
As we will see, its derivative with respect to $\beta$ has a minimum
which location is very close to $\beta_c$. Finally, the quantity $Q_v$
is defined as the squared ratio of the mean deviation and the standard
deviation (thus a relative of $P_W$ above) i.e.
\begin{equation}
  Q_v = \left\lbrack
  \frac{
    \langle | q_{\ell} - \langle q_{\ell}\rangle| \rangle^2
  }{
    \langle(q_{\ell}- \langle q_{\ell}\rangle)^2\rangle
  }\right\rbrack
  \label{defQv}
\end{equation}
Obviously $Q_v$ can only be found by storing the actual distributions
$Q(q_{\ell})$ for individual samples during simulation rather than
just the raw moments.

At high temperatures, i.e. $\beta \to 0$, for both bimodal and
Gaussian ISG interactions the $Q(q_{\ell})$ distributions become
symmetric, Gaussian, and centered on $q_{\ell}=0$, so $Q_{k}(0)=3,
Q_{s}(0)=0, Q_{w}(0)=0$. As $\beta$ is increased through $\beta_c$ the
$Q(q_{\ell})$ distributions become fat tailed and asymmetric, so
$Q_{k}(\beta,L)$ and $Q_{s}(\beta,L)$ show peaks in the region of
$\beta_c$.

The amplitudes of the peaks decrease with increasing $L$; this could
be called an "evanescent" critical phenomenon as it will disappear in
the thermodynamic limit. Allowing for a weak finite size correction
term, the positions of the maxima $\beta_{\mathrm{max}}$ for each set
of peaks tend towards $\beta_c$ with increasing $L$.

Physically, the peaks in the excess $Q$ kurtosis ("fat tailed"
distributions) and the $Q$ skewness near $\beta_c$ in ISGs must be
related to the build up of inhomogeneous temporary correlated spin
clusters around criticality. The data show that these clusters do not
produce a visible effect on the form of the $Q(q_{\ell})$ distribution
until $L$ is smaller than the thermodynamic correlation length
$\xi(\beta)$, which is related to the typical cluster size. For larger
$L$ the cluster effects average out in $Q(q_{\ell})$. Only when the
ratio $L/\xi(\beta)$ is smaller than some value so that a cluster can
englobe the entire sample do deviations from the Gaussian form
appear. As $\xi(\beta)$ diverges at $\beta_{c}$, the $Q$ kurtosis and
the $Q$ skewness will each tend to a peak for fixed $L$, and the peaks
can be expected to be situated exactly at $\beta_c$ in the large $L$
limit. Indeed analogous behavior can be observed in a pure Ising
ferromagnet, with excess $Q$ kurtosis and $Q$ skewness peak positions
tending towards $\beta_c$ with increasing $L$ \cite{lundow:12}.

The $Q_{\mathrm{var}}(\beta,L)$ and $Q_{w}(\beta,L)$ parameters are
closely related as $\lbrack\langle q_{\ell}\rangle(\beta, L)\rbrack$
becomes almost independent of $L$ at large $L$. We will show data for
$\log (Q_{\mathrm{var}}(\beta,L)-1)$ and $Q_{w}(\beta,L)$. Both of
these parameters have basically the form of a phenomenological
coupling, with curves for different $L$ intersecting at crossing
points $\beta_{\mathrm{cross}}(L,L^{'})$ which approach $\beta_c$ as
$L$ and $L^{'}$ are increased. The corrections to scaling for the two
parameters are slightly different.  As the inter-sample variability
for these parameters is much weaker in the region of $\beta_c$ than is
the equivalent variability for the phenomenological couplings based on
spin overlap distributions, they provide an accurate tool for
estimations of the critical temperature.

\section{Critical exponent estimates}

The thermodynamic limit spin glass susceptibility including the Wegner
confluent correction to scaling term \cite{wegner:72} is
\begin{equation}
  \chi(\tau) = C_{\chi}\tau^{-\gamma}(1+a_{\chi}\tau^{\theta} + b_{\chi}\tau + \cdots).
  \label{defchi}
\end{equation}
with the natural ISG scaling variable $\tau = 1-(\beta/\beta_{c})^2$
\cite{daboul:04,campbell:06}.  The analogous correlation length
expression is \cite{campbell:06}
\begin{equation}
  \xi(\tau)/\beta = C_{\xi}\tau^{-\nu}(1+a_{\nu}\tau^{\theta} + b_{\nu}\tau + \cdots).
  \label{defxi}
\end{equation}
with the same exponent $\theta$.  (Unfortunately ISG susceptibility
and correlation length data are most frequently analyzed using the
scaling variable $t = (T-T_{c})/T_{c} = \beta_{c}/\beta -1$ which is
inappropriate for ISGs except as an approximation in a narrow region
around the critical point.)  Once the value of $\beta_c$ can be
considered to be precisely determined, a further step is to make a
plot of the temperature dependent effective exponent $\gamma(\tau)$
\cite{butera:02,campbell:06} from the spin glass susceptibility data
$\chi(\beta)$, with the definition
\begin{equation}
  \gamma(\tau)= -\dd\log(\chi(\beta))/\dd\log(\tau)
  \label{defgamma}
\end{equation}
and the temperature dependent effective exponent $\nu(\tau)$ can be defined by
\begin{equation}
  \nu(\tau)= -\dd\log(\xi(\beta)/\beta)/\dd\log(\tau)
  \label{defnu}
\end{equation}
In a [hyper]cubic lattice, from the second term in the ISG HTSE the
limiting effective exponent at infinite temperature $\gamma(1)=
2d\beta_c^2$ exactly, where $2d$ is the number of nearest neighbors.
The ratio $\gamma(0)/\gamma(1)$ is directly related to the strength
and sign of the confluent correction coefficient $a_{\chi}$.  If the
remaining correction terms are negligible, then to leading order
\begin{equation}
  a_{\chi} \sim \frac{\gamma(0) - 2d\beta_c^2}{\theta}
  \label{achi}
\end{equation}
which gives a criterion for the strength and sign of the confluent
correction to scaling term (but not for the value of the exponent
$\theta$) once $\beta_{c}$ and $\gamma(0)$ have been estimated.

As long as $L \gg \xi(\beta)$ the finite size numerical data are $L$
independent and so are effectively in the thermodynamic limit infinite
size regime. The region where this condition holds for each $L$ can be
seen by inspection of $\gamma_{\mathrm{eff}}(\tau,L)$ and other
parameters such as $\nu_{\mathrm{eff}}(\tau,L)$.  $\chi(\beta)$ and
$d\chi/d\beta^{2}$ can be readily evaluated by direct summation of the
terms given in Ref.~\cite{daboul:04}, from small $\beta$ down to some
$\beta$ beyond which the contribution of further terms of greater than
$15$th order become non-negligible. These HTSE data can be used as a
check on the numerical data; in all cases agreement was good.  Then
from Eqn.~\eqref{defchi} one can plot
\begin{multline}
  \gamma(\tau) = \gamma_{c}- \dd\log\left(1+ a_{\chi}\tau^{\theta} + b_{\chi}\tau + \cdots\right)/\dd\log\tau
  \\ = \gamma_{c} - \left(a_{\chi}\theta\tau^{\theta} + b_{\chi} \tau + \cdots\right)
  \label{gameff}
\end{multline}
and
\begin{equation}
  \chi(\beta)\tau^{\gamma_{c}} = C_{\chi}\left(1+ a_{\chi}\tau^{\theta} + b_{\chi}\tau + \cdots\right)
  \label{chitaugam}
\end{equation}

Luckily, according to \cite{daboul:04} for the systems studied
$\theta$ is of the order of or a little greater than $1$, so the
leading and subleading corrections have about the same exponent. An
adequate analysis can be made using a joint effective correction term
with a single effective $\theta$. Then extrapolation to criticality at
$\tau=0$ can be made by first estimating the parameters $\gamma_c$
from the plot of $\gamma_{\mathrm{eff}}(\tau)$, Eqn.~\eqref{gameff}. Then, with
fixed $\beta_{c}$ and $\gamma(0)$ the scaled spin glass susceptibility
can be plotted in the form $\chi(\tau)\tau^{\gamma}$ against
$\tau^{\theta}$ , Eqn.~\eqref{chitaugam}, with the correction to
scaling exponent $\theta$ chosen such that the plot is a straight line
over the thermodynamic limit data region. The parameters $C_{\chi}$
and $a_{\chi}$ can be read off this plot.  All the parameters are
adjusted until the fits to both equations \eqref{chitaugam} and
\eqref{gameff} are optimised for the thermodynamic limit data. In
practice the fits lead to accurately determined values for
$\gamma_{c}$ and the other parameters, see
e.g. Fig.~\ref{fig:17}. This value is fully reliable under the unique
condition that $\beta_c$ has been correctly determined. An accurate
knowledge of $\beta_c$ is essential; there is a one-to-one
relationship between the estimate for the critical $\gamma(0)$ and the
value of $\beta_c$ taken to construct the plot. 


It can be noted that in the thermodynamic limit regime $L > \xi(\tau)$
there is "self-averaging", or in other words all individual ISG
samples of a system have the same properties and in particular the
same spin glass susceptibility. Thus in this regime there is no real
need to average over large numbers of samples to obtain accurate
measurements of the mean $\chi(\tau)$. In addition, equilibration in
the thermodynamic limit region is relatively rapid so measurements are
very reliable and not subject to equilibration difficulties. On the
contrary, in the regime near, at and beyond $\beta_c$ "lack of
self-averaging" sets in; the inter-sample variability is
important. The non-self-averaging parameters are size independent
\cite{hukushima:00,palassini:03} at and beyond $\beta_{c}$ so however
large the individual ISG samples they are all different from each
other; there is a wide distribution of values of the spin glass
susceptibility and other properties. The onset of a non-zero
variability is a spin glass criterion for an approach to the
transition temperature which obviously has no equivalent in pure
systems such as simple ferromagnets. Even in diluted ferromagnets the
non-self-averaging is non-zero in the thermodynamic limit only at
$\beta=\beta_{c}$.

\section{Numerical simulations}

For equilibration and measurements we used standard heat bath updating
(without parallel tempering) on randomly selected sites. The samples
(usually $64$) started off at infinite temperature and was then
gradually cooled before reaching their final designated
temperature. For temperatures near $T_c$ this means that each sample
went through at least $10^7$, sometimes $10^8$, sweeps before any
measurements took place.  Normally there were about 10 sweeps between
measurements, depending on temperature, maintaining on average $L^d$
spin flips between each measurement.  For each sample and temperature
we collected between $10^6$ and $10^7$ measurements depending on
lattice size.  The test for equilibration was discussed above.  It can
be noted that a sample with $L=8$ in dimension $5$ corresponds to as
many individual spins as a sample with $L=32$ in dimension $3$.

\section{Bimodal ISG in dimension five}

No detailed numerical simulation measurements have been reported
before for ISGs in dimension five.  However, the analysis of a $15$
term High Temperature Series Expansion (HTSE) calculation
\cite{daboul:04} gave the estimate $\beta_{c}^2 =0.154(3)$,
i.e. $\beta_c = 0.3925(35)$, with a critical exponent $\gamma =
1.95(15)$, for the bimodal ISG in $5$d.  We have re-analyzed the two
series in an unorthodox but transparent manner, Appendix I, and obtain
values for $\beta_c$ and $\gamma$ very close to the central values in
the original HTSE analysis but with additional information as to the
strength of the correction to scaling term in the two cases. It can be
seen that the correction to scaling is strong in the $5$d bimodal case
(and practically negligible in the $5$d Gaussian case).

\begin{figure}
  \includegraphics[width=3.in]{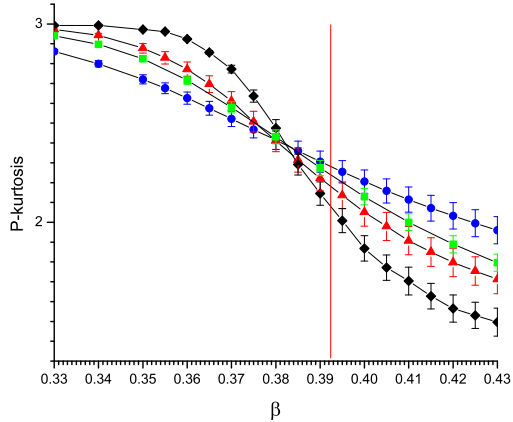}
  \caption{(Color online) The mean $P$ kurtosis Eqn.~\eqref{defPk} for
    $5$d bimodal interaction samples with $L=4$ (blue circles), $L=5$
    (green squares), $L=6$ (red triangles) and $L=8$ (black
    diamonds). The vertical red line corresponds to the HTSE
    $\beta_{c}$ central value \cite{daboul:04}.
  }\protect\label{fig:1}
\end{figure}

\begin{figure}
  \includegraphics[width=3.in]{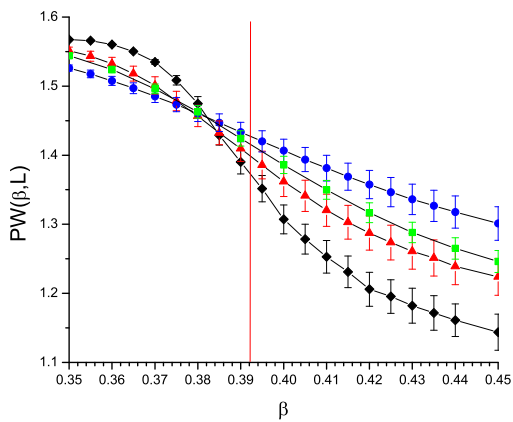}
  \caption{(Color online) The mean $P_W$ parameter Eqn.~\eqref{defPW} for
    the sets of $5$d bimodal interaction samples with $L=4, 5, 6, 8$
    (sizes coded as in Fig.~\ref{fig:1}). The vertical red line
    corresponds to the HTSE $\beta_{c}$ central
    value.}\protect\label{fig:2}
\end{figure}

\begin{figure}
  \includegraphics[width=3.in]{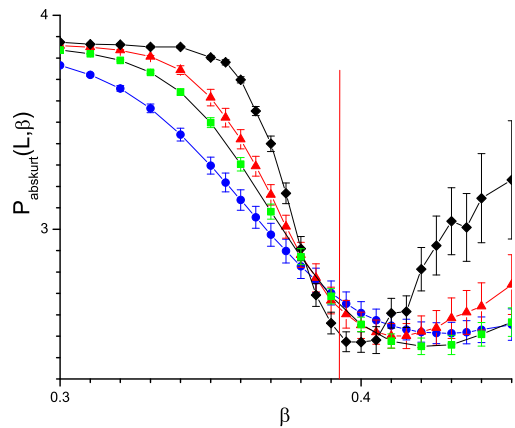}
  \caption{(Color online) The mean kurtosis of the absolute value spin
    overlap distribution, $P_{\mathrm{abskurt}}(\beta,L)$, Eqn.~\eqref{defPabk},
    for the sets of $5$d bimodal interaction samples $L=4, 5, 6, 8$
    (sizes coded as in Fig.~\ref{fig:1}). The vertical red line
    corresponds to the HTSE $\beta_{c}$ central
    value.}\protect\label{fig:3}
\end{figure}

\begin{figure}
  \includegraphics[width=3.in]{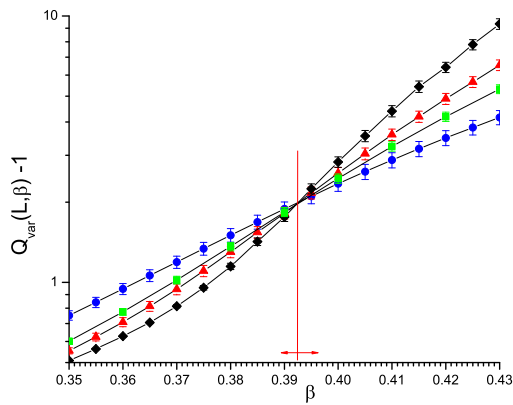}
  \caption{(Color online) The logarithm of the $Q$ variance
    Eqn.~\eqref{defQvar} minus $1$, $\log(Q_{\mathrm{var}}-1)(\beta,L)$, for the
    sets of $5$d bimodal interaction samples with $L=4, 5, 6, 8$
    (sizes coded as in Fig.~\ref{fig:1}). The vertical red line
    corresponds to the HTSE $\beta_{c}$ central value. The horizontal
    red line represents the HTSE $\beta_{c}$ range
    \cite{daboul:04}.}\protect\label{fig:4}
\end{figure}

\begin{figure}
  \includegraphics[width=3.in]{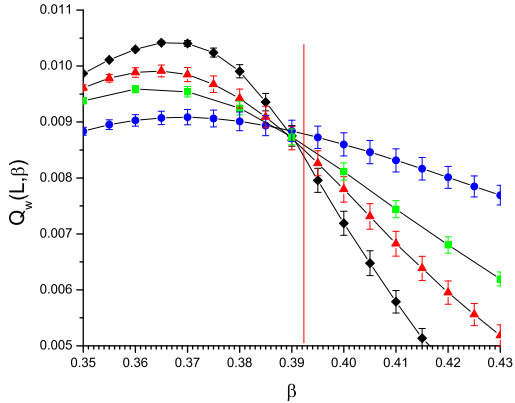}
  \caption{(Color online) The inverse normalized $Q$ variance
    $Q_{w}(\beta,L)$ Eqn.~\eqref{defQw}, for the sets of $5$d bimodal
    interaction samples with $L=4, 5, 6, 8$ (sizes coded as in
    Fig.~\ref{fig:1}). The vertical red line corresponds to the HTSE
    $\beta_{c}$ central value.}\protect\label{fig:5}
\end{figure}

\begin{figure}
  \includegraphics[width=3.in]{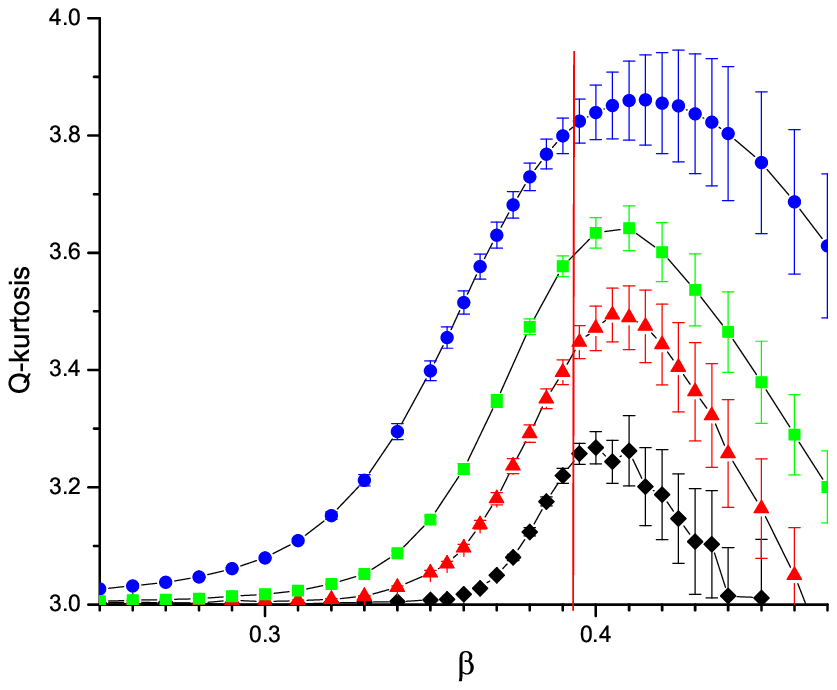}
  \caption{(Color online) The $Q$ kurtosis Eqn.~\eqref{defQk},
    $Q_{k}(\beta,L)$, for the sets of $5$d bimodal interaction samples
    with $L=4, 5, 6, 8$ (sizes coded as in Fig.~\ref{fig:1}). The
    vertical red line corresponds to the HTSE $\beta_{c}$ central
    value.}\protect\label{fig:6}
\end{figure}

\begin{figure}
  \includegraphics[width=3.in]{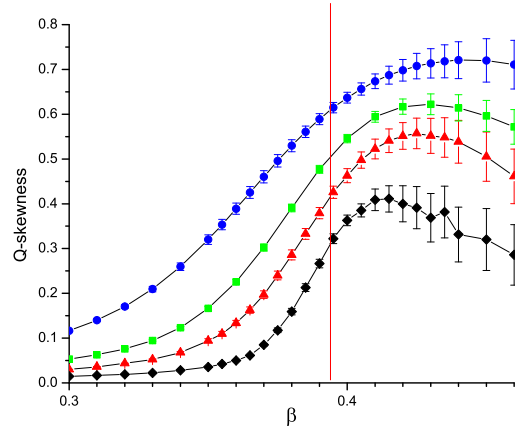}
  \caption{(Color online) The $Q$ skewness Eqn.~\eqref{defQsk},
    $Q_{s}(\beta,L)$, for the sets of $5$d bimodal interaction samples
    with $L=4, 5, 6, 8$ (sizes coded as in Fig.~\ref{fig:1}). The
    vertical red line corresponds to the HTSE $\beta_{c}$ central
    value.}\protect\label{fig:7}
\end{figure}

Numerical data derived from the various $P(q)(\beta,L)$ and
$Q(q_{\ell})(\beta,L)$ distributions, all taken in the same runs on
the same sets of samples for each $L$, are shown in Figures
\ref{fig:1} to \ref{fig:7}. The error bars correspond to inter-sample
variability for each particular parameter.

The spin overlap $P(q)$ based phenomenological couplings $P$ kurtosis
Eqn.~\eqref{defPk}, $P_W$ Eqn.~\eqref{defPW}, and the skewness of the
absolute $P$ distribution Eqn.~\eqref{defPabskew} show very similar
forms. Because of the strong dispersion of individual sample parameter
values, crossing points derived from the present data with a modest
number of samples scatter and cannot provide an accurate estimate for
$\beta_c$. The data are broadly compatible with the central HTSE
estimate but on their own do not provide anything like a critical
test, which for these parameters would require averaging over a much
larger number of samples.

With the present sets of of samples, the most useful spin overlap
parameter is the kurtosis of the absolute value distribution,
$P_{\mathrm{abskurt}}(\beta,L)$ Eqn.~\eqref{defPabk}. It shows a strong dip for
each $L$ in the region of the HTSE $\beta_c$ estimate. The center of
the dip as a function of temperature can be estimated quite accurately
for each $L$. When the central dip positions are plotted against $1/L$
and extrapolated to $1/L=0$, the data give an estimate $\beta_{c} =
0.392(3)$. This is the most precise estimate obtained from the spin
overlap distributions and is fully consistent with the HTSE central
value.

The link overlap parameters $Q_{w}(\beta,L)$ and
$\log(Q_{\mathrm{var}}-1)(\beta,L)$ show much smaller inter-sample variability
than the parameters based on spin overlap. By luck
$\log(Q_{\mathrm{var}}-1)(\beta,L)$ has negligible finite size correction to
scaling for the $5$d bimodal case; the curves for different $L$ all
intersect at the same $\beta_{\mathrm{cross}}(L,L^{'})= 0.392(1)$, which
coincides with the central value for $\beta_c$ from the HTSE estimate
$\beta_c = 0.3925(35)$ \cite{daboul:04}. This result both validates
the assumption that this link overlap parameter is a {\it bona fide}
phenomenological coupling, and improves the precision on the value of
the critical temperature. For $Q_{w}(\beta,L)$ there are weak finite
size corrections to scaling, but the intersection points
$\beta_{\mathrm{cross}}(L,L^{'})$ as functions of $L^{'}$ for fixed $L$ can be
extrapolated to infinite $L^{'}$ to obtain an estimate
$\beta_c=0.392(3)$ which is again in agreement with the HTSE value. In
addition the (negative) maximum in the derivative $\dd
Q_{w}(\beta,L)/\dd\beta$ deepens with increasing $L$ and its position
$\beta_{\mathrm{max}}(L)$ can with extrapolation also be used to
estimate $\beta_{c}$.  The intersection criterion and the maximum
slope criterion conveniently bracket the critical $\beta_{c}$ more and
more closely as the sizes are increased. From these $Q_{\mathrm{var}}$
and $Q_{w}$ numerical data alone one can thus derive a very precise
estimate $\beta_{c} = 0.392(1)$, entirely consistent with the HTSE
central estimate.  The $Q$ kurtosis $Q_{k}$ and $Q$ skewness $Q_{s}$ peak
positions are subject to finite size corrections but are also
consistent with this estimate for $\beta_{c}$.

Thus the $5$d bimodal ISG is a particularly favorable case to validate
the assumption that the link overlap parameters show strictly critical
forms in ISGs as they do in a ferromagnet \cite{lundow:12}.

\section{The Gaussian ISG in dimension 5}

The critical temperature was estimated from the HTSE analysis to
correspond to $\beta_{c}^2 =0.174, 0.176(3)$ or $0.177(3)$ according
to the different analysis techniques \cite{daboul:04}, i.e. $\beta_c =
0.420(3)$. The spin glass critical exponent was estimated by the HTSE
analysis to be $\gamma = 1.75(15)$.

\begin{figure}
  \includegraphics[width=3.in]{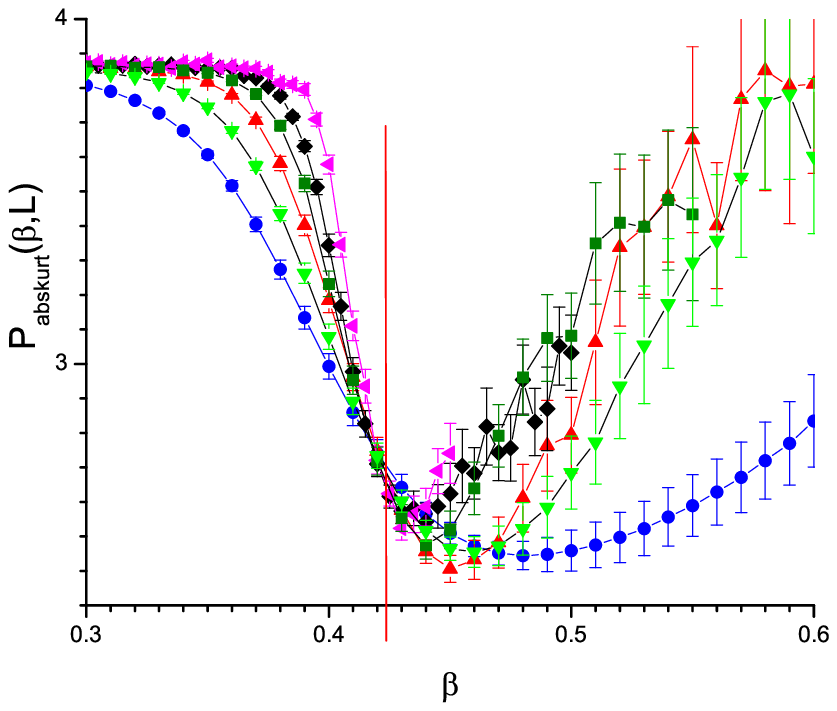}
  \caption{(Color online) The $P_{\mathrm{abskurt}}(\beta,L)$,
    Eqn.~\eqref{defPabk}, for the sets of $5$d Gaussian interaction
    samples with $L= 4, 5, 6, 7, 8, 10$ (blue circles, green inverted
    triangles, red triangles, olive squares, black diamonds, pink left
    triangles) . The vertical red line corresponds to the HTSE central
    value $\beta_{c}=0.421$.}\protect\label{fig:8}
\end{figure}

\begin{figure}
  \includegraphics[width=3.in]{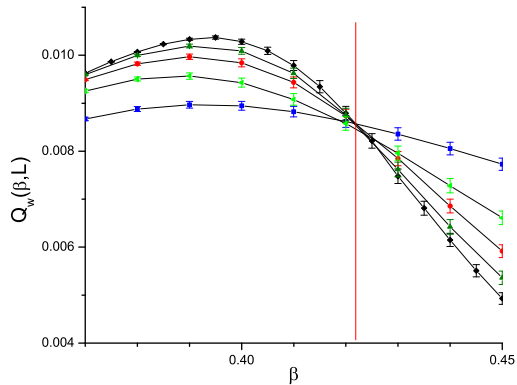}
  \caption{(Color online) The $Q_{w}(\beta,L)$ parameter of
    Eqn.~\eqref{defQw}, for the sets of $5$d Gaussian interaction samples
    with $L=4, 5, 6, 7, 8$ (sizes coded as in Fig.~\ref{fig:8}). The
    vertical red line corresponds to the HTSE central value
    $\beta_{c}=0.421$.}\protect\label{fig:9}
\end{figure}

\begin{figure}
  \includegraphics[width=3.in]{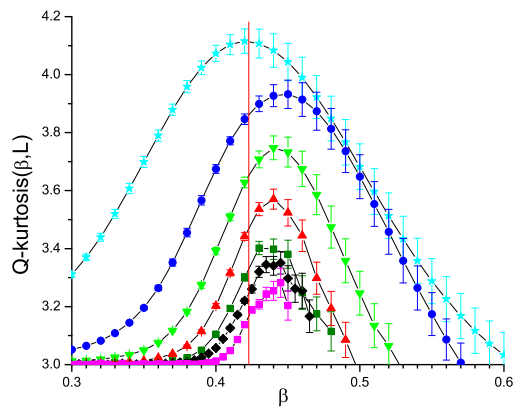}
  \caption{(Color online) The $Q$ kurtosis $Q_{k}(\beta,L)$
    Eqn.~\eqref{defQk}, for the sets of $5$d Gaussian interaction samples
    with $L=3, 4, 5, 6, 7, 8, 10$ (sizes coded as in
    Fig.~\ref{fig:8}). The vertical red line corresponds to the HTSE
    central value$\beta_{c}=0.421$.}\protect\label{fig:10}
\end{figure}

\begin{figure}
  \includegraphics[width=3.in]{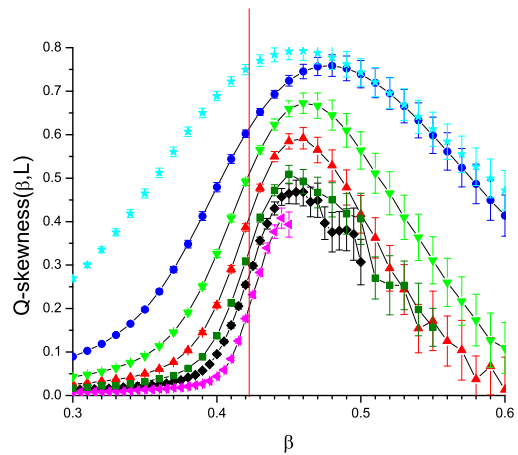}
  \caption{(Color online) The $Q$ skewness $Q_{s}(\beta,L)$
    Eqn.~\eqref{defQsk}, for the sets of $5$d Gaussian interaction
    samples with $L=3, 4, 5, 6, 7, 8, 10$ (cyan stars, blue circles,
    green inverted triangles, red triangles, olive squares, black
    diamonds, pink left triangles) . The vertical red line corresponds
    to the HTSE central value $\beta_{c}=0.421$.}\protect\label{fig:11}
\end{figure}

The data for spin and link overlap moments and moment ratios are shown
in Figs~\ref{fig:8} to \ref{fig:11}.  The general form for each
parameter is similar to that for the $5$d bimodal ISG, with the
appropriate $\beta_c = 0.421(2)$ being estimated from the absolute P
kurtosis and from $Q_{w}$ allowing for corrections to scaling, in
agreement with the central value from the HTSE analysis.

In the data for for the $Q$ kurtosis and the $Q$ skewness,
Fig.~\ref{fig:10} and Fig.~\ref{fig:11}, the peak positions are
tending towards $\beta_{c}$ with increasing $L$ more slowly than in
the bimodal case. This may be due to strong finite size corrections or
possibly a peculiarity of the Gaussian interaction distribution.

\section{The bimodal ISG in dimension 4}

From an analysis of HTSE data for the $4$d bimodal ISG, Daboul et al
\cite{daboul:04} estimate $\beta_{c}^2 = 0.26(2)$,
i.e. $\beta_{c}=0.51(2)$. (HTSE estimates in $4$d are intrinsically
less precise than in $5$d).  A critical temperature
$\beta_{c}=0.493(7)$ was estimated \cite{marinari:99} from simulation
measurements of high statistical accuracy to $L=10$ using the Binder
parameter crossing point criterion, but corrections to scaling were
not allowed for.  A further estimate is $\beta_{c} = 0.5025(25)$
\cite{bernardi:97} from unpublished Binder parameter data to $L=12$ by
A.P. Young. From extensive domain wall free energy measurements to
$L=10$ Hukushima gives an estimate $T_{c} =2.00(4)$
\cite{hukushima:99}, i.e. $\beta_c=0.50(1)$. In fact the raw data show
significant finite size corrections, which affect the extrapolated
estimate for the value of $\beta_c$ in the infinite size limit. This
can be seen clearly in the data shown in Fig.~$4$ of
Ref.~\cite{hukushima:99}; the crossing points
$T_{\mathrm{cross}}(L,L^{'})$ evolve regularly from the smallest sizes
to the largest sizes measured : $T_{\mathrm{cross}}(4,6) \sim 2.20,
T_{\mathrm{cross}}(6,8) \sim 2.06, T_{\mathrm{cross}}(8,10)\sim 2.00$.
By inspection, the infinite size limit crossing temperature must be
distinctly lower than $T = 2.00$, i.e. $\beta_c > 0.50$
\cite{hukushima}.

\begin{figure}
  \includegraphics[width=3.in]{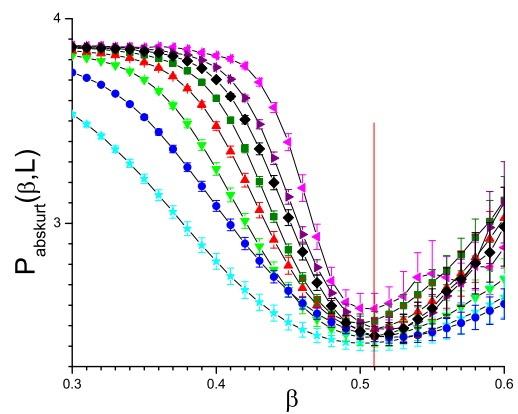}
  \caption{(Color online) The absolute $P$ kurtosis
    $P_{\mathrm{abskurt}}(\beta,L)$, Eqn.~\eqref{defPabk}, for the sets of $4$d
    bimodal interaction samples with $L=3, 4, 5, 6, 7, 8, 9, 12$ (cyan
    stars, blue circles, green inverted triangles, red triangles,
    olive squares, black diamonds, purple right triangles, pink left
    triangles). The vertical red line corresponds to the HTSE central
    value $\beta_{c}=0.51$}\protect\label{fig:12}
\end{figure}

\begin{figure}
  \includegraphics[width=3.in]{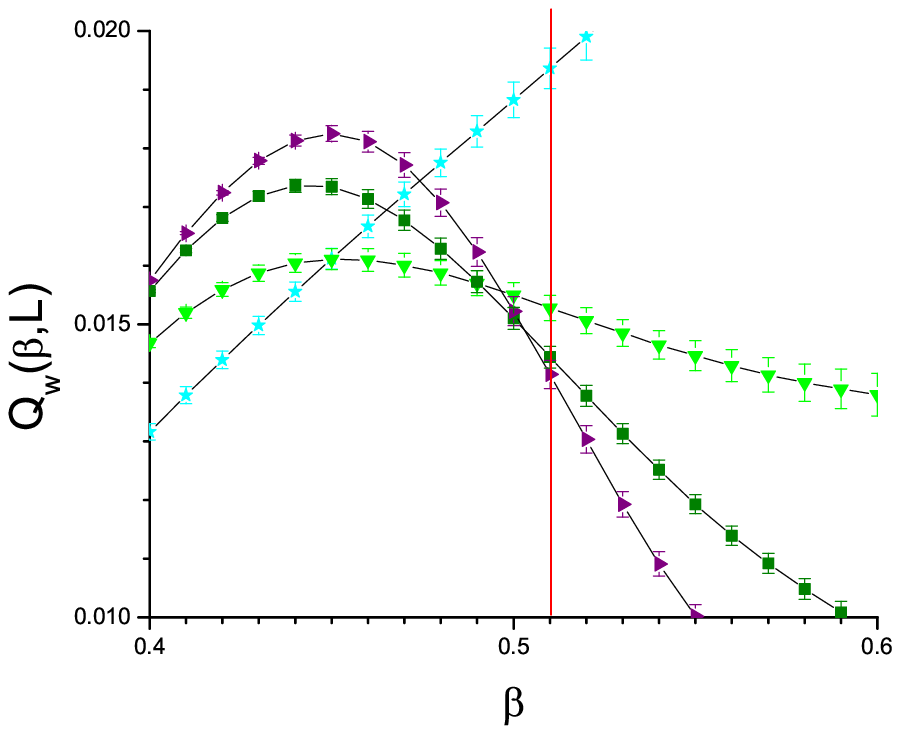}
  \caption{(Color online) The parameter $Q_{w}(\beta,L)$,
    Eqn.~\eqref{defQw}, for odd $L$ sets of $4$d bimodal interaction
    samples with $L=3, 5, 7, 9$ (size coding as in
    Fig.~\ref{fig:12}). The vertical red line corresponds to the HTSE
    central value $\beta_{c}=0.51$}\protect\label{fig:13}
\end{figure}

\begin{figure}
  \includegraphics[width=3.in]{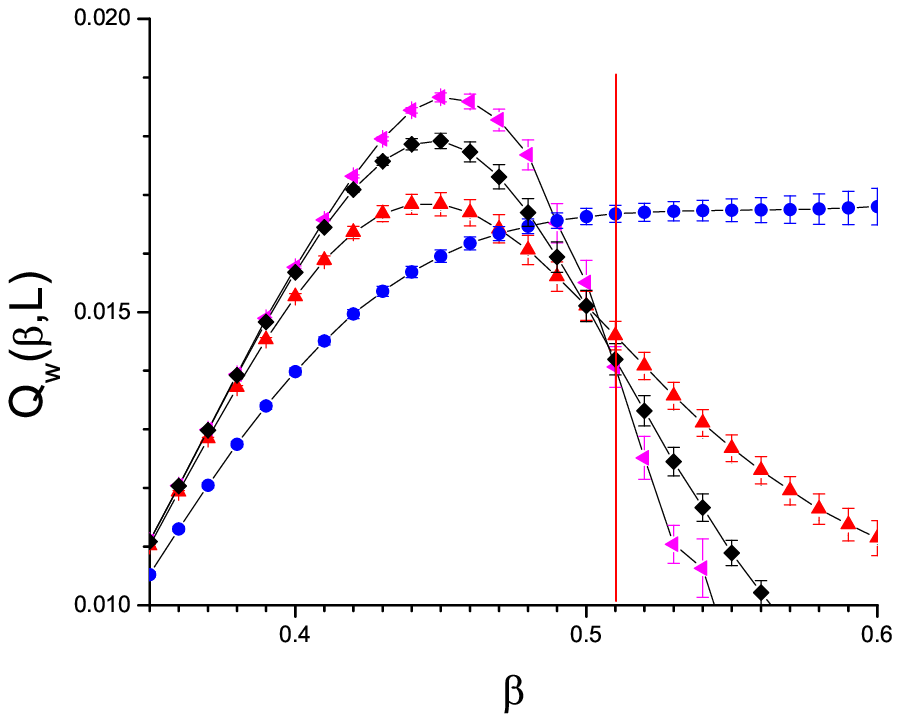}
  \caption{(Color online) The parameter $Q_{w}(\beta,L)$,
    Eqn.~\eqref{defQw}, for even $L$ sets of $4$d bimodal interaction
    samples with $L=4, 6, 8, 12$ (size coding as in
    Fig.~\ref{fig:12}). The vertical red line corresponds to the HTSE
    central value $\beta_{c}=0.51$}\protect\label{fig:14}
\end{figure}

\begin{figure}
  \includegraphics[width=3.in]{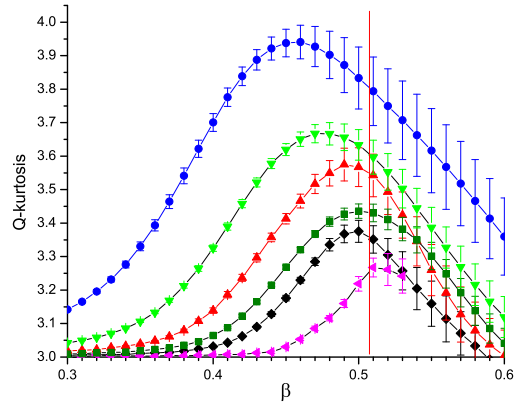}
  \caption{(Color online) The $Q$ kurtosis $Q_{k}(\beta,L)$,
    Eqn.~\eqref{defQk}, for sets of $4$d bimodal interaction samples
    with $L=4, 5, 6, 7, 8$ and $12$ (size coding as in
    Fig.~\ref{fig:12}). The vertical red line corresponds to the HTSE
    central value $\beta_{c}=0.51$}\protect\label{fig:15}
\end{figure}

\begin{figure}
  \includegraphics[width=3.in]{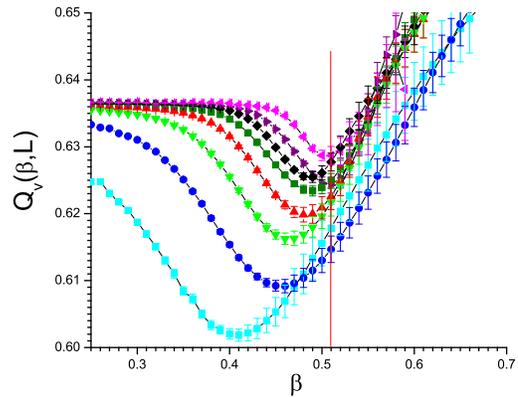}
  \caption{(Color online) The $Q$ deviation ratio $Q_{v}(\beta,L)$,
    Eqn.~\eqref{defQv}, for sets of $4$d bimodal interaction samples
    with $L=3, 4, 5, 6, 7, 8, 9$ and $12$ (size coding as in
    Fig.~\ref{fig:12}). The vertical red line corresponds to the HTSE
    central value $\beta_{c}=0.51$.  }\protect\label{fig:15v}
\end{figure}

\begin{figure}
  \includegraphics[width=3.in]{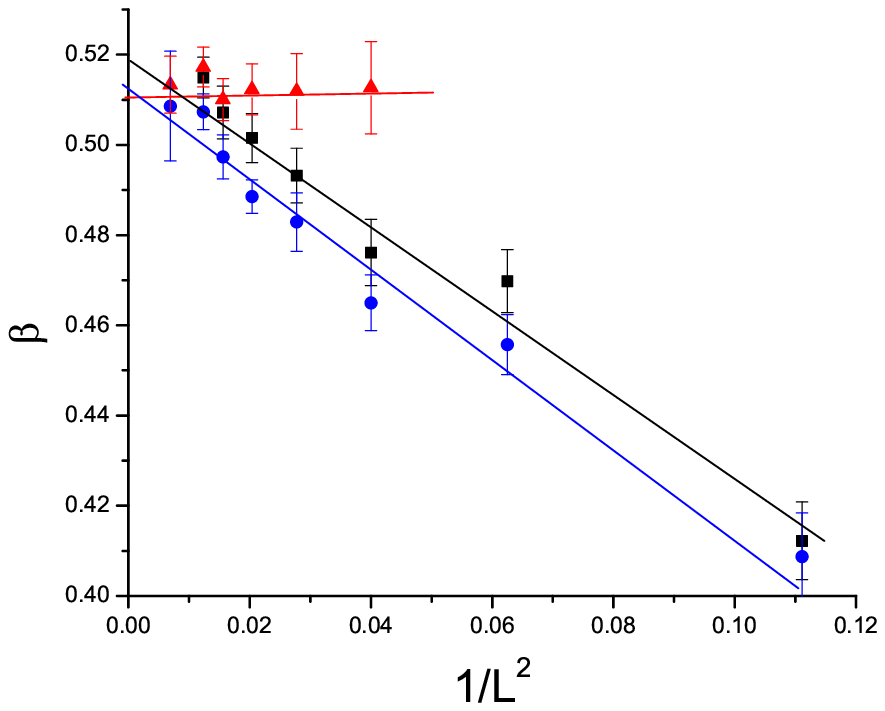}
  \caption{(Color online) The $Q$ kurtosis $Q_{k}(\beta,L)$ peak
    temperature $\beta_{\mathrm{max}}(L)$ (black squares), the $Q_v$
    minimum location (blue circles) and the $\dd Q_w/\dd\beta$ minimum
    location (red triangles) against $1/L^{2}$ for sets of $4$d
    bimodal interaction samples with $L=3, 4, 5, 6, 7, 8, 9$.  See
    Figs.~\ref{fig:13}, \ref{fig:14}, \ref{fig:15} and \ref{fig:15v}
    (derivative). The extrapolation to infinite $L$ gives
    $\beta_{c}=0.510(5)$.  Error bars were obtained from statistical
    resampling of half the data set.  }\protect\label{fig:16}
\end{figure}

Simulations were carried out on sets of $64$ samples of size $L=3, 4,
5, 6, 7, 8, 9$ and $12$. As in $5$d the phenomenological couplings
based on the spin overlap were strongly affected by the inter-sample
variability; much larger sets would have been needed to obtain
crossing point data for these parameters of similar statistical
precision as in the earlier results for the Binder parameter.  However,
the position of $\beta_{\mathrm{dip}}$, the minimum of the dip in
$P_{\mathrm{abskurt}}$, Fig~\ref{fig:12}, is independent of $L$ at
$\beta_{\mathrm{dip}} = 0.506(3)$ to within the statistical precision.


For the link overlap parameter $Q_{w}(\beta,L)$, Fig.~\ref{fig:13} and
Fig.~\ref{fig:14}, the inter-sample variability is much weaker than
for the spin overlap phenomenological coupling parameters, so the
crossing points for successive $L$ are better determined, but there
are both finite size corrections with the crossing points evolving
towards larger $\beta_{\mathrm{cross}}(L,L^{'})$ with increasing
$L,L^{'}$, and odd-even effects in $L$. 


Extrapolating to $1/L = 0$ the $\beta_{\mathrm{cross}}(L − 1,L + 1)$
values for crossing points between $Q_w(\beta,L − 1)$ and $Q_w(\beta,L
+ 1)$ in Figs.~\ref{fig:13} and \ref{fig:14} leads to the estimate
$\beta_{c} = 0.515(5)$ from this criterion.  The deviation ratio
$Q_v$, Fig.~\ref{fig:15v}, also shows weak sample variability with a
distinct minimum approaching $\beta_c$.  The $Q$ kurtosis
$Q_{k}(\beta,L)$, Fig.~\ref{fig:15}, and $Q$ skewness $Q_{s}(\beta,L)$
peak positions evolve with increasing $L$ towards limiting $\beta$
values for $1/L=0$ which are consistent with the estimate from
$dQ_{w}/d\beta$, Fig.~\ref{fig:16}. It can be concluded that
$\beta_{c} = 0.510(5)$ in full agreement with the central HTSE
estimate \cite{daboul:04} and with the previous numerical measurements
once finite size corrections are fully allowed for.

\section{The Gaussian ISG in dimension 4}

High precision simulation measurements have been published for the
$4$d Gaussian ISG, and for a $4$d diluted bimodal ISG
\cite{jorg:08}. The critical temperature for the $4$d Gaussian ISG was
estimated from Binder parameter and correlation length ratio
measurements to be $\beta_c =0.554(3)$ in full agreement with earlier
simulation estimates $0.555(3)$ \cite{parisi:96,ney:98} and with the
HTSE estimate $\beta_{c}^2 =0.314(4)$, i.e. $\beta_{c}=0.560(3)$.

\begin{figure}
  \includegraphics[width=3.in]{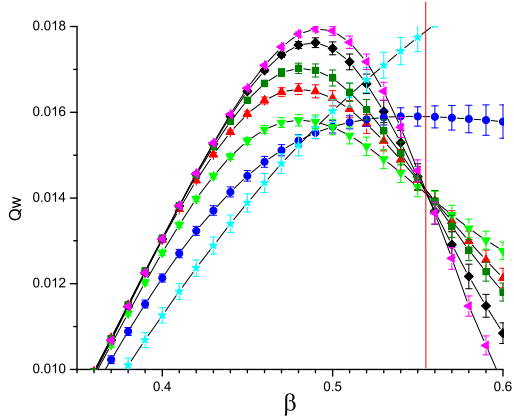}
  \caption{(Color online) The link overlap parameter $Q_w(\beta,L)$
    for the $4$d Gaussian, $L=3,4,5,6,7,8,10$ (size coding as in
    Fig.~\ref{fig:12})}\protect\label{fig:16b}
\end{figure}

\begin{figure}
  \includegraphics[width=3.in]{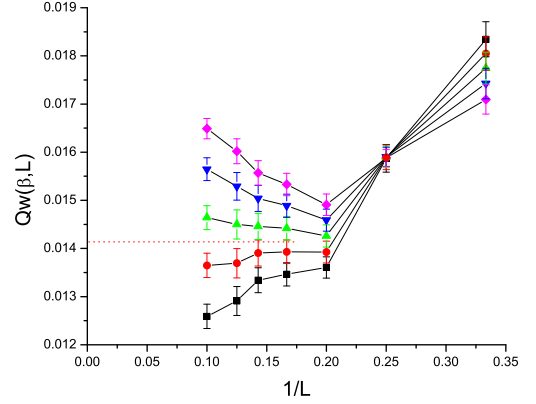}
  \caption{(Color online) The link overlap parameter $Q_w(\beta,L)$ for
    the $4$d Gaussian against $1/L$ for fixed $\beta$.  Pink diamonds,
    blue inverted triangles, green triangles, red circles, black
    squares for $\beta = 0.53, 0.54, 0.55, 0.56, 0.57$. The horizontal
    dashed line indicates the critical behavior in the large $L$
    limit.}\protect\label{fig:16c}
\end{figure}

Link overlap data for the parameter $Q_w$ measured for $64$ samples at
each size are shown in Fig.~\ref{fig:16b} and \ref{fig:16c}. It can be
seen that there are systematic finite size effects for the positions
of the intersections between curves for different $L$, but these
corrections have already become almost negligible by the largest sizes
studied here as can be seen in Fig.~\ref{fig:16c}.  Even with the
modest number of samples in these simulations, the link overlap $Q_w$
data provide an accurate independent estimate $\beta_c=0.554(2)$ which
confirms the value of Ref.~\cite{jorg:08}.

\section{The critical exponent $\gamma$}

Daboul {\it et al} \cite{daboul:04} concluded that in each dimension
the HTSE critical bimodal and Gaussian $\gamma$ values for the
different interaction distributions which they studied were compatible
with universality in ISGs to within the uncertainties of the HTSE
analysis. However, their error bars for each $\gamma$ value were
relatively large, as they did not have access to simulation data which
supplement the HTSE calculations and which refine both the $\beta_{c}$
and the $\gamma$ estimates.

\begin{figure}
  \includegraphics[width=3.in]{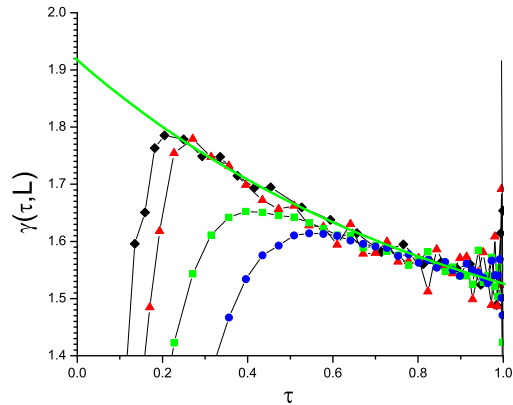}
  \caption{(Color online) The effective exponent $\gamma(\tau)$
    Eqn.~\eqref{defgamma} for the bimodal $5$d ISG samples with $L=4, 5,
    6, 8$ (size coding as in Fig.~\ref{fig:1}), assuming
    $\beta_{c}=0.3925$. The extrapolation to $\tau=0$ gives $\gamma=
    1.92(5)$.
  }\protect\label{fig:17}
\end{figure}

\begin{figure}
  \includegraphics[width=3.in]{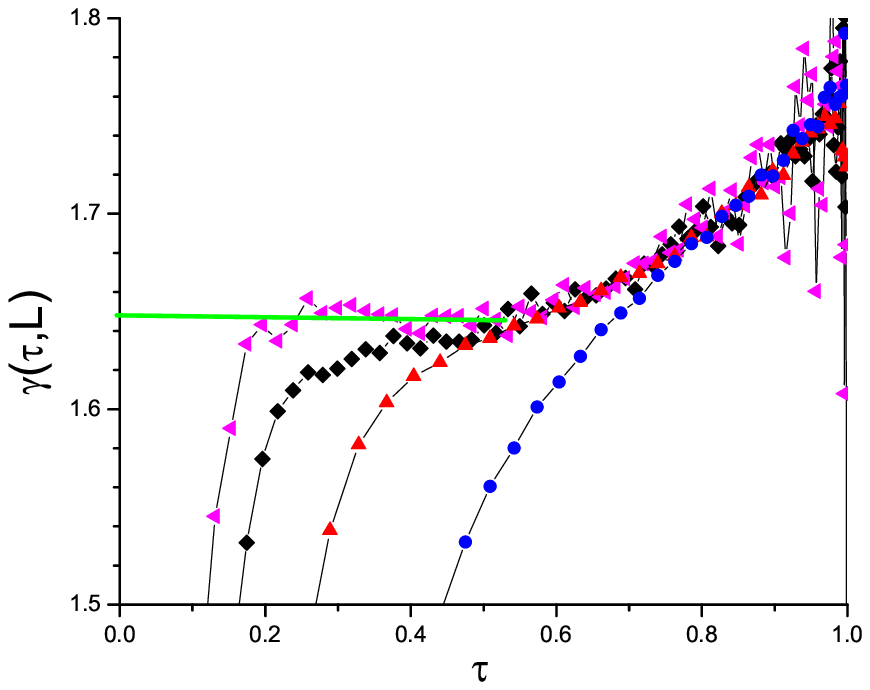}
  \caption{(Color online) The effective exponent $\gamma(\tau)$
    Eqn.~\eqref{defgamma} for Gaussian $5$d ISG samples with $L=4, 6, 8,
    10$ (size coding as in Fig.~\ref{fig:8}), assuming
    $\beta_{c}=0.421$. The extrapolation to $\tau=0$ gives $\gamma=
    1.66(3)$.}\protect\label{fig:18}
\end{figure}

\begin{figure}
  \includegraphics[width=3.in]{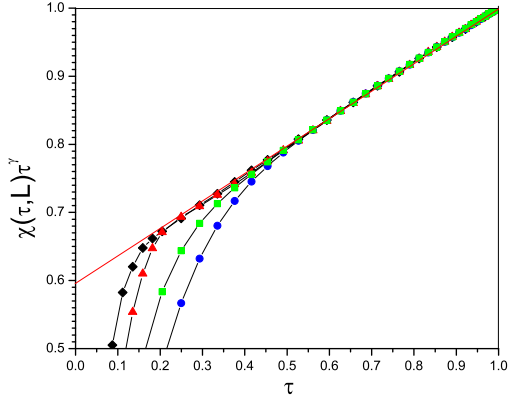}
  \caption{(Color online) The reduced susceptibility
    $\chi(\tau)\tau^{\gamma}$ for the bimodal $5$d ISG samples $L=4,
    5, 6, 8$ (size coding as in Fig.~\ref{fig:8}), assuming
    $\beta_{c}=0.3925$, $\gamma=1.92$. The overall thermodynamic limit
    fit is $\chi(\tau)= 0.59\tau^{-1.92}[1+0.69\tau]$.
  }\protect\label{fig:19}
\end{figure}

\begin{figure}
  \includegraphics[width=3.in]{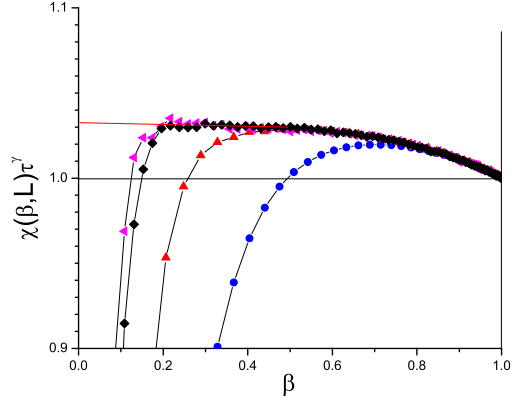}
  \caption{(Color online) The reduced susceptibility
    $\chi(\tau)\tau^{\gamma}$ for the Gaussian $5$d ISG samples $L=4,
    6, 8, 10$ (size coding as in Fig.~\ref{fig:8}), assuming
    $\beta_{c}=0.421$, $\gamma=1.66$. The overall thermodynamic limit
    fit is $\chi(\tau)= 1.03\tau^{-1.66}[1+\cdots]$.
  }\protect\label{fig:20}
\end{figure}

The effective $\gamma(\tau,L)$ values are defined by
Eqn.~\eqref{defgamma}, see the "Critical exponent estimates"
section. For the $5$d bimodal and $5$d Gaussian systems
$\gamma(\tau,L)$ is shown in Fig.~\ref{fig:17} and Fig.~\ref{fig:18},
with in each case $\tau= 1-(\beta/\beta_{c})^2$, the data being
plotted using the optimal values $\beta_{c}(\mathrm{bimodal})=0.392$
and $\beta_{c}(\mathrm{Gaussian})=0.421$ as quoted above.  See also
Figs.~\ref{fig:19} and \ref{fig:20} for a plots of the reduced
susceptibility in these cases with the assumed values of $\beta_c$ and
critical exponents.


It can be noted that in $5$d a sample with $L=8$ is "large" in the
sense that for this size $\chi(\tau,L)$ remains in the thermodynamic
limit until $\tau \sim 0.2$. To approach $\beta/\beta_{c} = 1$ equally
closely in $3$d would require samples with $L \sim 100$. The
difference is a consequence of the much higher value of the exponent
$\nu$ in dimension $3$, which is not far from the lower critical
dimension where $\nu$ diverges.

For the $5$d bimodal ISG the plot of the effective $\gamma(\tau)$
shows a critical limit for the extrapolated thermodynamic limit regime
$\gamma(0) = 1.92(5)$, and a strong correction to scaling with
exponent $\theta \sim 1$. With this value of $\gamma$ in hand
$\chi(\tau)\tau^{\gamma}$ can be plotted as a function of $\tau$,
Fig.~\ref{fig:19}, which can be fitted by $\chi(\tau)\tau^{\gamma}=
0.59(1+0.69\tau)$. So finally for the $5$d bimodal ISG, in
Eqn.~\eqref{defchi} $\gamma=1.92(5), \theta \sim 1, C_{\chi}=0.59,
a_{\chi}=0.69$. As $\theta \sim 1$ the second correction term in
Eqn.~\eqref{defchi} cannot be distinguished from the leading term. The
fit curve for $\gamma_{\mathrm{eff}}(\tau)$ in Fig.~\ref{fig:20} is the
derivative of $\chi(\tau)$ with the same parameters. The parameter set
including the sign and the approximate strength of $a_{\chi}$ are
consistent with those which can be estimated entirely independently
from the HTSE data (see Appendix I). The critical $\gamma$ and
$\theta$ values are consistent with but more accurate than the HTSE
value $\gamma = 1.95(15)$ and $\theta \sim 1.1$ of Daboul {\it et al}
\cite{daboul:04}.

For the $5$d Gaussian ISG the plot of the effective $\gamma(\tau)$
shows a critical limit $\gamma(0) = 1.66(3), C_{\chi}=1.03$. The
Daboul {\it et al} HTSE estimate is $\gamma= 1.75(15)$. The correction
to scaling is very weak except for a high order term in the region far
from criticality, so no estimate can be made for $\theta$. Again the
HTSE data, as analysed in Appendix I, are in complete agreement with
all these conclusions : a very similar critical temperature, a very
similar critical exponent, and the same weak correction to critical
scaling.

The principal conclusion which can be drawn from the $5$d results,
both from the simulations and from the HTSE data analyses, is that the
critical exponent $\gamma =1.92(5)$ for the bimodal ISG is
significantly higher than the critical $\gamma = 1.66(3)$ for the
Gaussian ISG.

\begin{figure}
  \includegraphics[width=3.in]{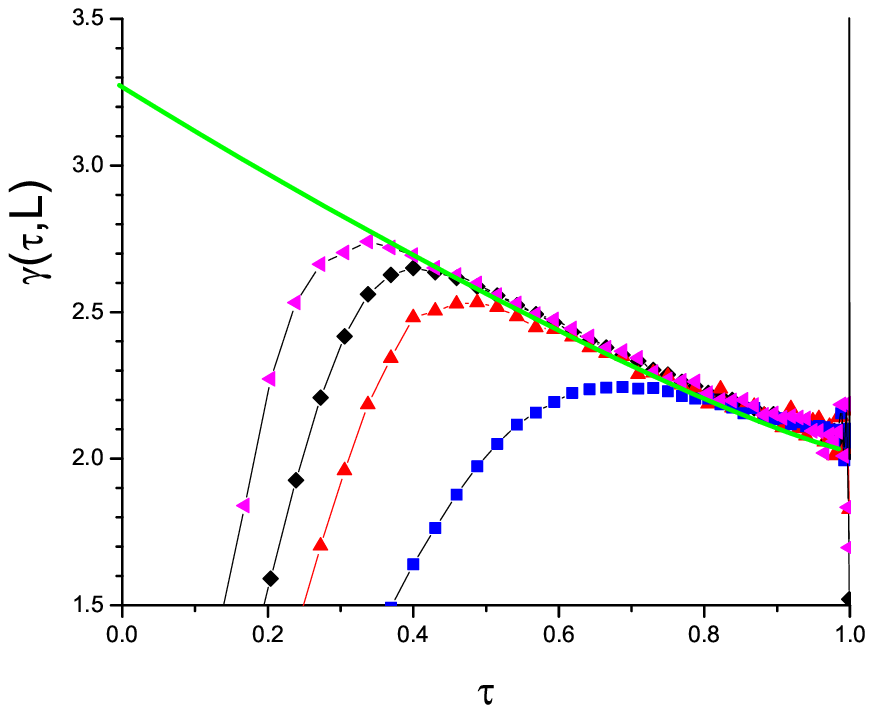}
  \caption{(Color online) The effective exponent $\gamma(\tau)$
    Eqn.~\eqref{defgamma} for the bimodal $4$d ISG samples with $L=4, 6,
    8, 12$ (size coding as in Fig.~\ref{fig:12}), assuming
    $\beta_{c}=0.51$. The extrapolation to $\tau=0$ gives $\gamma=
    3.2(1)$}\protect\label{fig:21}
\end{figure}

\begin{figure}
  \includegraphics[width=3.in]{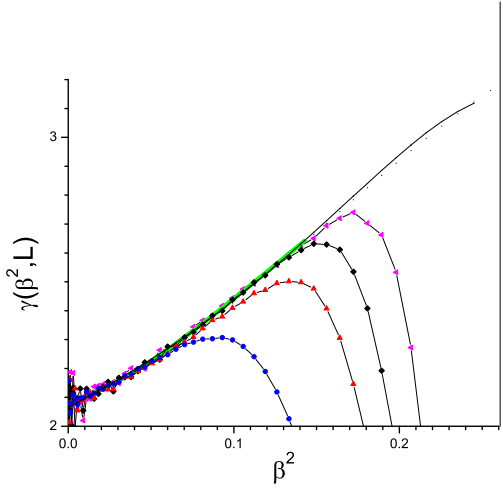}
  \caption{(Color online) The effective $\gamma(\beta^2,L)$ presented
    in a different way for the bimodal $4$d ISG, assuming the HTSE
    critical temperature squared $\beta_{c}^{2} = 0.26$. The
    simulation data $L=4,6,8,12$ have the same color coding as in
    Fig.~\ref{fig:12}. The thick green curve shows the explicitly
    summed HTSE data from the series given in \cite{daboul:04}. The
    thin black curve is the optimal polynomial fit to the
    thermodynamic limit regime data (series and numerical)
    extrapolated to $\beta_{c}^2$.  The right hand side vertical line
    is at $\beta_{c}^{2}=0.26$.  }\protect\label{fig:21b}
\end{figure}

\begin{figure}
  \includegraphics[width=3.in]{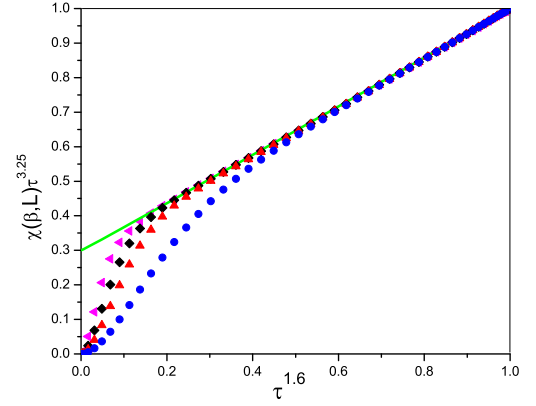}
  \caption{(Color online) The reduced susceptibility
    $\chi(\tau)\tau^{\gamma}$ for the bimodal $4$d ISG samples $L=4,
    6, 8, 12$ (size coding as in Fig.~\ref{fig:12}), assuming
    $\beta_{c}=0.51$, $\gamma=3.2$, $\theta = 1.6$. The overall
    thermodynamic limit fit is $\chi(\tau)= 0.30\tau^{-3.2}[1+
      2.3\tau^{1.6}]$.
  }\protect\label{fig:22}
\end{figure}

The $4$d bimodal $\gamma(\tau)$ plots assuming $\beta_c =0.51$ are
shown in Figs.~\ref{fig:21} and \ref{fig:21b}. Extrapolating the
thermodynamic limit regime curve to $\tau = 0$ gives a critical
exponent estimate $\gamma(0) = 3.25(10)$ where the error bar
corresponds principally to the residual uncertainty in
$\beta_{c}$. The fit parameters to $\chi(\tau)\tau^{\gamma}$,
Fig.~\ref{fig:22}, are $C_{\chi} = 0.30$, $a_{\chi} = 2.3$ and $\theta
\sim 1.6$, so there is a very strong correction to scaling.


The estimate quoted from the HTSE analysis \cite{daboul:04} was
$\gamma = 2.5(3)$, for the same central value of $\beta_c$ as in the
present work. We do not understand this. The raw HTSE susceptibility
data are in excellent agreement with the numerical data (as they
should be) and show a $\gamma_{\mathrm{eff}}(\tau)$ which is
increasing rapidly as criticality is approached;
$\gamma_{\mathrm{eff}}(\tau)$ is already greater than $2.5$ by
$\tau=0.5$.
 
A re-analysis of unpublished $4$d bimodal $\chi(\beta)$ and
$\xi(\beta)$ data of Ref.~\cite{hukushima}, are in full agreement with
the present data as far as $\chi(\beta)$ is concerned.  Fixing
$\beta_c = 0.51$ and defining $\nu_{\mathrm{eff}}(\tau) =
\dd\log(\xi(\tau)/\beta)/\dd\log(\tau)$~\cite{campbell:06}, leads to
an estimate for the correlation length critical exponent $\nu =
1.30(5)$.

High precision simulation measurements have been made of the $4$d
Gaussian ISG and of a $4$d diluted bimodal ISG \cite{jorg:08}. The
critical temperature for the $4$d Gaussian ISG was estimated from
Binder parameter and correlation length ratio measurements to be
$\beta_c =0.554(3)$ in full agreement with earlier simulation
estimates $0.555(3)$ \cite{parisi:96,ney:98} and with the HTSE
estimate $\beta_{c}^2 =0.314(4)$ i.e. $\beta_{c}=0.560(3)$. 

\begin{figure}
  \includegraphics[width=3.in]{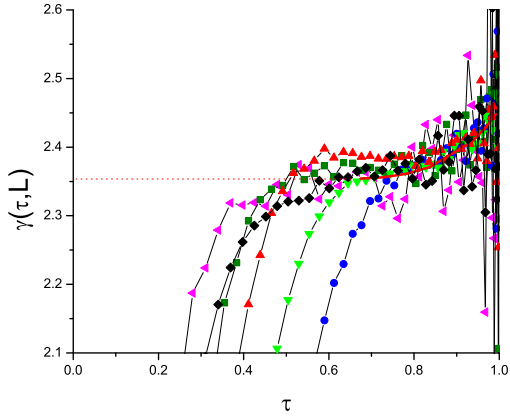}
  \caption{(Color online) The effective exponent $\gamma(\tau)$
    Eqn.~\eqref{defgamma} for the Gaussian $4$d ISG samples with $L=4,
    5, 6, 7, 8, 10$ (size coding as in Fig.~\ref{fig:12}), assuming
    $\beta_{c}=0.554$. The extrapolation to $\tau=0$ gives $\gamma=
    2.35(2)$.The thick red curve shows the explicitly summed HTSE data
    from the series given in \cite{daboul:04}.}\protect\label{fig:21c}
\end{figure}

For the Gaussian and the diluted bimodal ISG \cite{jorg:08} the
critical exponents were estimated to be $\eta = -0.275(25)$ and
$\nu=1.02(2)$ so $\gamma = (2-\eta)\nu = 2.32(8)$, and $\eta =
-0.275(25), \nu=1.025(15)$ so $\gamma = 2.33(6)$ respectively. The
simulation data showed that critical finite size corrections to
scaling are weak. This is consistent with the criterion for $a_{\chi}$
given above. With the Gaussian critical parameters : $a_{\chi} \sim
\gamma - z\beta_{c}^2 =2.32(8)-2.45(2)=-0.13(10)$; the Wegner
susceptibility correction to scaling amplitude will be small.
Simulation and HTSE data for $\gamma_{\mathrm{eff}}(\tau)$ assuming
$\beta_c =0.554$ are shown in Fig.~\ref{fig:21c}; it can be seen that
the corrections to scaling are indeed small, and by extrapolation to
$\tau=0$ we find a critical $\gamma = 2.35(2)$ in full agreement with
Ref.~\cite{jorg:08}.  This can taken as a validation of the
methodology used in the present work.

Naturally it was concluded in Ref.~\cite{jorg:08} that as diluted
bimodal and Gaussian $4$d ISGs have the same critical exponents to
within the statistical uncertainties, universality is
confirmed. However, the critical $\gamma = 3.25(10)$ estimated above
for the undiluted $4$d bimodal ISG is significantly higher than the
$\gamma \sim 2.33$ values estimated for the Gaussian and the diluted
bimodal ISGs.  It happens that at the particular diluted bimodal bond
concentration studied in Ref.~\cite{jorg:08}, $p=0.35$, the kurtosis
of the bond distribution is $1/0.35$, so almost exactly equal to the
Gaussian distribution kurtosis which is $3$. It is tempting to
speculate that there could be a universality rule for ISGs such that
at fixed dimension, the exponents depend on the kurtosis of the
interaction distribution, just as the critical temperature of an ISG
in each dimension is a function of the kurtosis of the interaction
distribution \cite{campbell:05}.  For the moment we have not studied
the ISGs in dimension $3$.


\section{Conclusion}

The moments and moment ratios of the link overlap distributions in
ISGs show well defined critical properties, analogous to those
observed for the link overlap distribution moments in a simple
ferromagnet \cite{lundow:12}. The inter-sample variability of the link
overlap parameters in the ISGs is weaker than that of the spin overlap
parameters, so link overlap critical measurements, even with modest
numbers of independent samples, are intrinsically more precise than
spin overlap measurements. With larger numbers of samples, similar to
those used in earlier simulation studies (e.g. Ref.~\cite{jorg:08}),
and with negligible supplementary computational cost, extremely
accurate $\beta_{c}$ values could be obtained from link overlap
parameters.

Link overlap critical data have been used here to supplement spin
overlap data and HTSE analyses in order to obtain accurate estimates
for the ordering temperatures of ISGs in dimensions $4$ and $5$. We
have also introduced a useful spin overlap dimensionless parameter,
the absolute P distribution kurtosis $P_{\mathrm{abskurt}}$,
Eqn.~\eqref{defPabk}, which has not been previously studied.  The
$\beta_{c}$ values estimated from these simulations are all in
excellent agreement with the entirely independent central estimates
from HTSE analyses \cite{daboul:04}, but the estimates for the
critical exponent $\gamma$ are more accurate. It should again be
underlined that to obtain precise estimates of critical exponents it
is essential to first establish reliable values for the critical
temperatures.  Once the ordering temperatures in hand, the effective
critical exponents $\gamma(\tau)$ can be readily and reliably
estimated from the appropriate derivative of the spin glass
susceptibility simulation data, Eqn.~\eqref{defgamma}, which can be
extrapolated to $\tau=0$ to obtain $\gamma$.

The present critical $\gamma$ estimates -- 5d bimodal $\gamma =
1.92(5)$, 5d Gaussian $\gamma = 1.66(3)$, 4d bimodal $\gamma =
3.2(1)$, 4d Gaussian $\gamma = 2.35(2)$ -- can be compared with values
obtained from analyses of the HTSE coefficients \cite{daboul:04} for
both $5$d and $4$d, and compared with published simulation results
\cite{jorg:08} for the $4$d Gaussian case.  The simulation data show
that both in $5$d and $4$d the critical $\gamma$ value for the bimodal
ISG is significantly higher than the critical $\gamma$ values for the
Gaussian ISG.


The well established universality rules which apply to standard second
order transitions are that systems having the same spatial and spin
dimensionalities all have identical critical exponents. From the
present data it can be concluded empirically that different, more
complicated, rules govern universality classes in Ising spin
glasses. It should be remembered that ISG transitions are
qualitatively very different from standard second order
transitions. For an Ising ferromagnet in the regime below the Curie
temperature there are just two mirror image families of spin up and
spin down states. For an ISG the non-self-averaging behavior means
that at and beyond $\beta_{c}$ even in the thermodynamic limit each
individual sample has different properties, in particular a different
spin overlap distribution and so a different spin glass
susceptibility. It is not obvious that the powerful renormalization
arguments which are so effective in standard transitions can be
applied in the same manner in this context.  It would be of interest
to explore from fundamental principles which relevant parameters
determine critical exponents in the spin glass family of transitions.

There are rare known cases of non-universality, such as the eight vertex
model \cite{baxter:71} and the Ashkin-Teller model
\cite{kadanoff:77}, which in the language of conformal invariance are
all related to field theoretical models with central charge $c \ge 1$
(see e.g. \cite{sowinski:13}). However, it is not clear to us if this
is relevant to the ISG situation.

\section{Acknowledgements} 
We are very grateful to K. Hukushima for comments and communication of
unpublished data.  The computations were performed on resources
provided by the Swedish National Infrastructure for Computing (SNIC)
at High Performance Computing Center North (HPC2N).

\section{Appendix}

HTSE calculations in ISGs \cite{daboul:04} produce a set of terms for
the spin glass susceptibility of the form
\begin{equation}
  \chi(\beta^2) = 1 + a_{1}\beta^2 +a_{2}\beta^4 + \cdots
  \label{htse}
\end{equation}
Each coefficient $a_{i}$ is exact, but the series is in practice
limited in length. The published ISG calculations have $15$ terms.

This series can be compared to the mathematical identity
\begin{multline}
  (1-x)^{-\gamma}= 1 + \gamma x + \frac{\gamma (1+\gamma)}{2} x^2 \\
  + \frac{\gamma (1+\gamma) (2+\gamma)}{6} x^3 + \cdots
  \label{series}
\end{multline}
for which the ratio of successive coefficients $a_{n}$ is
\begin{equation}
  \frac{a_{n+1}}{a_{n}} = 1+\frac{\gamma-1}{n+1}.
  \label{ratio1}
\end{equation}
In the simplest case of a physical system with a critical spin glass
susceptibility
$\chi(\beta^2)=C_{\chi}(1-\beta^2/\beta_{c}^2)^{-\gamma}$, the ratio
of successive terms $a_{n+1}/a_{n}$ in Eqn.~\eqref{htse} would be
$(1/\beta_{c}^2)(1 + (\gamma-1)/(n+1))$.  This suggests a graphical
analysis in terms of a plot of this ratio against $1/(n+1)$, which is
indeed a traditional technique for analyzing HTSE coefficients
\cite{fisher:64,butera:02}.  There are two complications. One is the
Wegner confluent correction to scaling \cite{wegner:72}; the critical
susceptibility is
\begin{equation}
  \chi(\tau)=C_{\chi}(\tau)^{-\gamma}\left(1+ a_{\chi}\tau^{\theta} +\cdots\right)
\label{wegner}
\end{equation}
where $\tau = 1-\beta^2/\beta_{c}^2$. The ratios become (see \cite{butera:02})
\begin{equation}
  \frac{a_{n+1}}{a_{n}} =
  \frac{1}{\beta_{c}^2} \left(1+\frac{\gamma-1}{n+1}
  - \frac{a_{\chi} \theta \Gamma(\gamma)}{\Gamma(\gamma-\theta) (n+1)^{1+\theta}}\right)
  \label{ratio2}
\end{equation}
so the initial slope is still $(\gamma-1)/\beta_{c}^2$ but there is a
higher order term, which means that the plot of the ratio against
$1/(n+1)$ becomes curved. This behavior is simply a reflection of the
true temperature dependence of $\chi(\tau)$.

\begin{figure}
  \includegraphics[width=3.in]{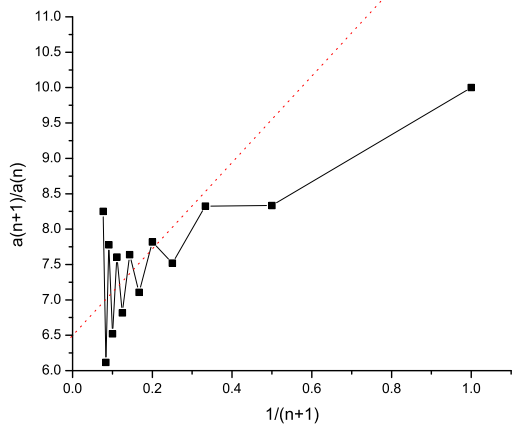}
  \caption{(Color online) The raw coefficient ratios $a(n+1)/a(n)$ in
    the HTSE susceptibility series for the bimodal ISG in dimension 5
    ($a(n)$ data from \cite{daboul:04}).
  }\protect\label{fig:23}
\end{figure}

\begin{figure}
  \includegraphics[width=3.in]{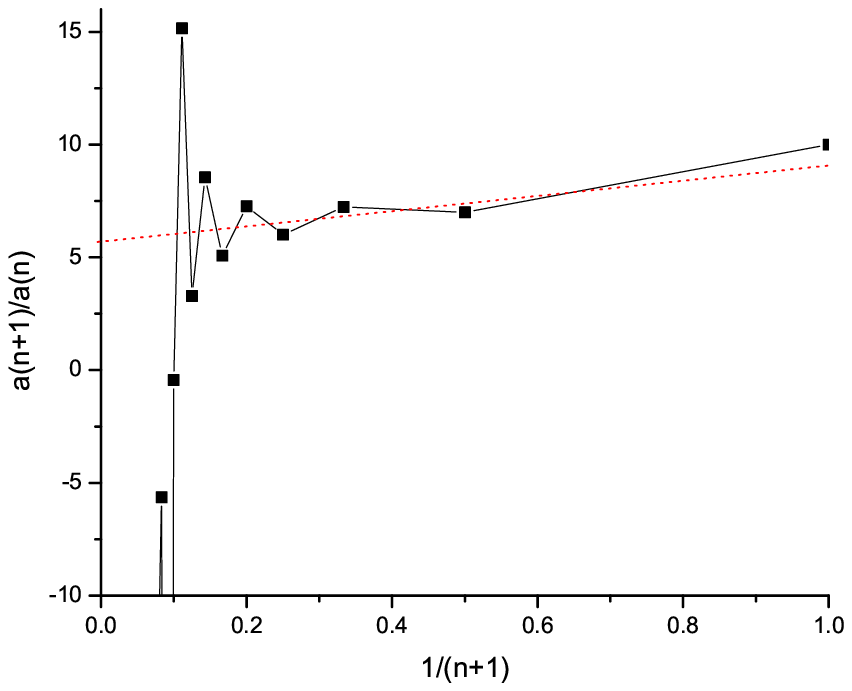}
  \caption{(Color online)The raw coefficient ratios $a(n+1)/a(n)$ in
    the HTSE susceptibility series for the Gaussian ISG in dimension 5
    ($a(n)$ data from \cite{daboul:04}).
  }\protect\label{fig:24}
\end{figure}

The second more annoying complication consists of series of
"parasitic" terms with alternating signs which arise from the presence
of anti-ferromagnetic poles \cite{daboul:04}. Although when summed to
infinite $n$ they give a zero or negligible contribution to the true
susceptibility $\chi(\tau)$, in the ISG case they can lead to dramatic
oscillations in the ratios $a_{n+1}/a_{n}$, Fig.~\ref{fig:23} and
Fig.~\ref{fig:24}. The series can nevertheless be analyzed, at least
in dimension $4$d and above, using the Pad\'e approximant technique,
accompanied by methods known as $M1$ and $M2$ \cite{daboul:04}.

\begin{figure}
  \includegraphics[width=3.in]{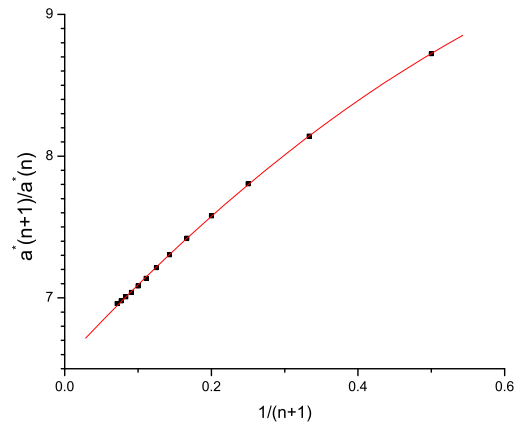}
  \caption{(Color online) The regrouped coefficient ratios
    $a^{*}(n+1)/a^{*}(n)$ in the HTSE susceptibility series for the
    bimodal ISG in dimension 5 for $\beta^2
    =9$.
  }\protect\label{fig:25}
\end{figure}

\begin{figure}
  \includegraphics[width=3.in]{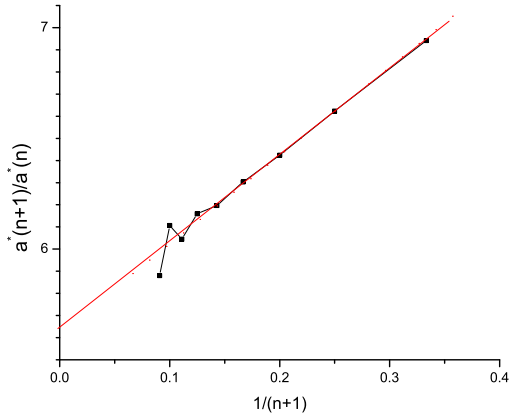}
  \caption{(Color online) The regrouped coefficient ratios
    $a^{*}(n+1)/a^{*}(n)$ in the HTSE susceptibility series for the
    Gaussian ISG in dimension 5 for $\beta^2
    =10$.
  }\protect\label{fig:26}
\end{figure}

An unorthodox but transparent variant on the graphical method is the
following.  Suppose the initial series for $\chi(\beta)$ at some
temperature $\beta < \beta_{c}$, terminating with term $n$, is written
\begin{equation}
\chi(n,x) = 1 +a_{1}x +a_{2}x^2 +a_{3}x^3+\cdots + a_{n}x^n.
\end{equation}
where $x = \beta^2$.
Then terms can be regrouped
\begin{multline}
  \chi^{*}(n+1,x) =
  1 +\left(\frac{3a_{1}}{4}+\frac{a_{2}x}{4}\right) x \\ 
  +\left(\frac{a_{1}}{4x} +\frac{a_{2}}{2}+\frac{a_{3} x}{4}\right) x^2\\ 
  +\left(\frac{a_{2}}{4x}+\frac{a_{3}}{2}+\frac{a_{4} x}{4}\right) x^3+\cdots\\ 
  +\left(\frac{a_{n}}{4x} +\frac{a_{n+1}}{2}+\frac{a_{n+2} x}{4}\right) x^{n+1} \\ 
  = 1 + a^{*}_{1}x + \cdots + a^{*}_{n+1}x^{n+1} \\ 
  = 1 + a_1 x +\cdots + a_n x^n \\
  + \frac{3 a_{n+1}}{4} x^{n+1} + \frac{a_{n+2}}{4} x^{n+2}
  \label{chistar}
\end{multline}
The two sums $\chi(n,x)$ and $\chi^{*}(n+1,x)$ are identical up to the
$n$th term. The series do not terminate in exactly the same place, but
this is irrelevant as the aim is to extrapolate to infinite $n$ so as
to obtain the true total $\chi(x)$.  The essential point is that the
sets of ratios of the successive coefficients $R^{*}(n+1) =
a^{*}_{n+1}/a^{*}_{n}$ in the regrouped $\chi^{*}$ series now evolve
smoothly with $1/(n+1)$, as can be seen in Figs.~\ref{fig:25} and
\ref{fig:26}, and can be readily extrapolated to infinite $n$. This
was certainly not the case for the raw series. The sum of the
extrapolated regrouped terms can be considered a very good estimate of
the true $\chi(x)$. It turns out that for the $5$d bimodal and
Gaussian ISGs the set of regrouped term ratios are very insensitive to
the choice of the trial $\beta^2$, and so the fit parameters for a
representative $\beta^2$ provide good estimates for the true critical
physical parameters. A simple fit $R^{*}(1/(n+1))= A
+B(1/(n+1)+C(1/(n+1)^2)$ provides estimates of the intercept
$A=1/\beta_c^2$, the initial slope $B=(\gamma-1)/\beta_c^2$ and the
strength of the Wegner correction to scaling $C = -
a_{\chi}\theta\Gamma(\gamma)/(\Gamma(\gamma-\theta)\beta_{c}^2)$ (For
the fit we have assumed for convenience $\theta \sim 1$
\cite{daboul:04} but other values can be chosen for $\theta$ ).  From
Fig.~\ref{fig:25}, $\beta_{c} =0.3905, \gamma= 1.85$ and $a_{\chi}\sim
2.0$ for the bimodal $5$d ISG, and from Fig.~\ref{fig:26}, $\beta_{c}
=0.420, \gamma= 1.68$ and $a_{\chi} \sim 0$ for the Gaussian $5$d
ISG. The values of the critical temperatures and the critical
exponents $\gamma$ are in excellent agreement with but appear to be
more accurate than the central values from the much more sophisticated
analysis of Daboul {\it et al} \cite{daboul:04}. In addition, this
method provides an estimate of the strength of the Wegner correction
term, which was not explicitly cited as a result of the analysis in
Ref.~\cite{daboul:04}.

It is important to underline that this technique is an analysis of the
exact HTSE coefficients and so is entirely independent of the
simulation data. Nevertheless the method again leads to a value for
the bimodal critical exponent $\gamma \sim 1.85$ which is quite different
from the Gaussian critical exponent $\gamma \sim 1.65$, in full
agreement with the conclusions drawn from the simulation data.

In Ref.~ \cite{daboul:04} the estimations of the critical exponents
$\gamma$ in dimensions $7$ and $8$, above the upper critical
dimension, are quoted as being greater than the exact theoretical
value $\gamma = 1$, which is suprising. The data can be reconciled
with theory if there are strong correction terms. Thus the explicitly
calculated $\chi(\tau)$ data points can be fitted by $\chi(\tau) =
2.0\tau^{-1}[1-0.5\tau^{0.24}]$ in dimension $7$ and by $\chi(\tau) =
1.6\tau^{-1}[1-0.375\tau^{0.30}]$ in dimension $8$. It can be
remembered that there are corrections to scaling above the upper
critical dimension in the pure ferromagnetic Ising model
\cite{berche:08}.

\end{document}